\documentstyle[spie,psfig,epsfig]{article} 

\title{Realizing 3D Spectral Imaging in the Far-Infrared: FIFI LS} 

\author{L. W. Looney\supit{a}, N. Geis\supit{a}, R. Genzel\supit{a}, 
W. K. Park\supit{a}, A. Poglitsch\supit{a},  W. Raab\supit{a}, \\
D. Rosenthal\supit{a}, A. Urban\supit{a}, and T. Henning\supit{b}
\skiplinehalf 
\supit{a}Max-Planck-Institut f\"ur Extraterrestrische Physik (MPE),
Postfach 1603, \\
D-85740 Garching, Germany\\
\supit{b}Astronomisches Institut Universit\"at Jena,
Schillerg\"a\ss chen 3,\\
D-07745 Jena, Germany 
}

\authorinfo{email: lwl@mpe.mpg.de; http://fifi-ls.mpe-garching.mpg.de}

\def\arcsec{\hbox{$^{\prime\prime}$}}
\def\arcdeg{\hbox{$^\circ$}}
 
\begin{document} 
  \maketitle 

\begin{abstract}

We present a progress report on the design and construction of the
Field-Imaging Far-Infrared Line Spectrometer (FIFI LS) for the SOFIA
airborne observatory.  
The design of the instrument is driven by the goal of maximizing
observing efficiency, especially for observations of faint, 
extragalactic objects.
Thus, FIFI LS utilizes an integral field technique
that uses slicer mirrors to optically re-arrange the two-dimensional
field into a single slit for a long slit spectrometer.
Effectively, a 5 $\times$ 5 pixel spatial field of view is imaged
to a 25 $\times$ 1 pixel slit and dispersed to a 
25 $\times$ 16 pixel, two-dimensional detector array, 
providing diffraction-limited spatial and spectral multiplexing.
In this manner, the instrument employs two parallel, medium 
resolution (R $\sim$2000) 
grating spectrometers for simultaneous observations in two bands: 
a short wavelength band (42 to 110 $\mu$m) and a long wavelength band
(110 to 210 $\mu$m).
Overall, for each of the 25 spatial pixels, the instrument
can cover a velocity range of $\sim$1500 km/s around selected 
far-infrared lines with an  estimated sensitivity of 
$2\times 10^{-15}$ W Hz$^{1/2}$ per pixel.
This arrangement provides good spectral coverage with
high responsivity.
{\bf This paper does not include Figures due to astro-ph size limitations.
Please download entire file at http://fifi-ls.mpe-garching.mpg.de/fifils.ps.gz.}

\end{abstract}


\keywords{Integral Field Imaging, Spectrometer, Far-Infrared, grating, optical slicer, FIR, FIFI LS, FIFI, SOFIA}

\section{INTRODUCTION}
\label{sect:intro}  

Astronomers will soon have access to unprecedented spatial
resolution and sensitivity in the far-infrared with the 
Stratospheric Observatory For Infrared Astronomy (SOFIA).
Far-infrared astronomical observations, which are impossible from 
the ground due to water absorption, are necessary to
understand fully a number of important astronomical problems
and issues.
Many astrophysical conditions require far-infrared probes because
the areas of interest are mostly inaccessible at other wavelengths 
due to severe extinction from
interstellar dust or the physics of interest is only manifest at far-infrared
wavelengths. In particular, far-infrared spectroscopy, pioneered and
developed on SOFIA's predecessor the Kuiper Airborne Observatory (KAO)
and greatly extended with the Infrared Space Observatory (ISO), 
will provide an injection of important data into astrophysical issues.

Building upon the success of our previous imaging Fabry-Perot far-infrared
spectrometer (FIFI\cite{fifi1,fifi2}) that was designed for the KAO,
we are developing a successor instrument for SOFIA: the Field-Imaging
Far-Infrared Line Spectrometer (FIFI LS\cite{us1,us2}).  
FIFI LS will utilize integral field spectral imaging in two 
wavelength bands: 42 to 110 $\mu$m and 110 to 210 $\mu$m.
This will allow the instrument to simultaneously
obtain dual wavelength band, spectral imaging of a 5 $\times$ 5 pixel field,
without scanning a Fabry-Perot or multiple pointings with
a long-slit spectrometer.
Thus, FIFI LS is
designed as a ``major-step'' forward and will take advantage of the
unique benefits that SOFIA offers.

With the increased sensitivity and resolution provided by SOFIA, the 
main scientific targets for FIFI LS will include the detailed
morphological studies of: (1) the heating and cooling of galaxies, (2) star
formation and the interstellar matter under low-metalicity conditions,
as found in dwarf galaxies, (3) active galactic nuclei and their
environment, (4) merging and interacting galaxies, and
(5) large surveys of nearby galaxies.
To reach our scientific goals, very high observing sensitivities and 
efficiencies are essential, requiring a compromise of 
spectral resolution; however, for
the science objectives listed above, a comparably low spectral resolution
(R $\sim$2000) is more than sufficient.
Overall, FIFI LS on SOFIA will be more sensitive than the 
ISO Long-Wavelength Spectrometer and have
much higher spatial resolution and mapping capabilities. 
As a future option, an extension of the instrument to the 25-42 $\mu$m 
range is planned upon availability of the Si:Sb detector arrays 
developed for SIRTF. 

\section{Instrument Design} 

\subsection{Spectrometer Concept}

FIFI LS achieves 2-dimensional spatial mapping and simultaneous spectral
multiplexing by optically slicing, or re-arranging, the 2D field of view
onto a single slit, which is then dispersed via a standard long-slit
spectrometer.  This type of ``optical slicer'' was originally devised
for laboratory spectroscopy\cite{lab} and later successfully implemented
for near-infrared astronomy\cite{3d}, but this is the first time that
this technique has been applied in the far-infrared.  Specifically,
FIFI LS has a 5 $\times$ 5 pixel field of view 
which is sliced into
five individual slitlets, re-arranged into a continuous 25 $\times$
1 pixel slit, and finally, fed into a grating spectrometer and dispersed
onto a 2D 16 $\times$ 25 pixel detector array.

One of the other unique aspects of FIFI LS is its dual-channel nature.
By using a dichroic beamsplitter in the entrance optics and two optical slicers,
FIFI LS covers the diffraction-limited field of view 
simultaneously in two wavelength bands:
a short wavelength band 42 to 110 $\mu$m and a long wavelength band 110
to 210 $\mu$m.
Fig. 1 shows how the dual channel integral field
system is realized in more detail as a projection of
the focal plane onto the two detector arrays with
diffraction-limited optics.
On the right side of the figure is the long wavelength channel
of the spectrometer which has a pixel scale of 14$\arcsec$
per pixel.
On the left side of the figure is the short wavelength channel
which has a pixel scale of 7$\arcsec$ per pixel.
This of course implies that the shorter wavelength channel has a smaller
field of view, as demonstrated in Fig. 1.
Overall, this scheme ensures that for all spatial elements in the field, 
spectra are observed simultaneously in the two bands, thereby
increasing observing efficency.

\begin{figure}
\begin{center}
\begin{tabular}{c}
\end{tabular}
\end{center}
\caption[example]{
Focal plane projections of the long and short wavelength
channels of FIFI LS, showing the instantaneous field of view
for both channels, the re-arrangement of the
5 $\times$ 5 field of view onto a 25 $\times$ 1 slit,
and the spectral projection of the
integral field onto the 16 $\times$ 25 detector arrays.
}
\end{figure} 

\subsection{Optical Design}

The geometrical layout, optimization, and analysis of the entire
spectrometer optics was carried out with the optical design software
ZEEMAX-EE Ver. 8.0. The two spectrometers were designed to yield 
diffraction-limited image quality: the geometrical spot diameters are small
compared to the diffraction disk diameter of a point source imaged by
the telescope.  A 3-D solid model of the complete 
FIFI LS optics is shown in Fig. 2, viewed from the top; all optical 
components are cooled to cryogenic temperatures.

\begin{figure}
\begin{center}
\begin{tabular}{c}
\end{tabular}
\end{center}
\caption{
Solid model of the FIFI LS optical system, viewed from
the top.
The IR light enters from the left and is split by a diachronic
into the long and short spectrometers--- the upper is
for short $\lambda$ and the lower is for long $\lambda$.
The 16 $\times$ 25 detector arrays are the biconic surfaces,
second from the right.
}
\end{figure}

A rotating K-mirror assembly (hidden on the left in Fig. 2) at liquid nitrogen 
temperature compensates
for rotation of the field of view during long integration times. Two
chopped calibrator sources with temperatures close to liquid nitrogen can
be switched into the beam for internal calibration and flat-fielding at
irradiation levels close to that of the SOFIA telescope background. The
calibrator sources are located near a pupil image so that the light
paths are equivalent for internal calibration and observation.
All optics after the calibration system are
at liquid helium temperature.
After the calibration optics, the beam enters the 
spectrograph through a
cold Lyot stop (not shown in Fig. 2) to suppress light diffracted by the
entrance aperture. A far-infrared dichroic beam splitter feeds both the
long and short wavelength spectrometer. In the short wavelength branch,
re-imaging optics double the image size on the image slicer.

\begin{figure}
\begin{center}
\begin{tabular}{c}
\end{tabular}
\end{center}
\caption{
A 3D model of the optical slicer assembly.
Light enters the slicer system from the upper-left field
mirror, and is deflected down toward the lowest surface, the slicer mirrors.
The five slicer mirrors re-image the pupil onto the 5 separated
capture mirrors, which is then focused onto the slit mirrors
that are at the center of the image, and final into the
spectrometer.
In the background is a grating and collimator mirror.
}
\end{figure}
 
In Figure 3, the optical slicer is shown in more detail.
The light enters into the slicer system from the flat
mirror (upper-left in Fig. 3), then continues down toward the slicer mirrors
(the cube-like object near the left bottom of Fig. 3).
These 5 mirrors in the mirror stack form the image slicer,
which acts as a field mirror creating a pupil image on one of 
the 5 re-imaging capture mirrors (the five mirrors near the top-middle). 
The capture mirrors re-arrange the images of the 5 slices along one,
slightly curved line on the slit-mirror,
near the center of the image. 
Working in combination, these three mirrors,
the slicer, capture, and slit mirrors, 
perform the slicing and re-arranging of the
field of view onto the entrance slit for a grating spectrograph.
In addition, the three mirror system re-aligns the pupil of
each slit, so that the virtual pupils of each slice coincide.
In this case, we do not use simple flat mirrors for the slicing
process, as is usually done in near-infrared instruments, 
because of the larger A$\times \Omega$ product (area
times the beam solid angle); 
curved mirrors allow for a much more compact slicer assembly design.

The FIFI LS spectrometers use slightly off-axis (0.5$\arcdeg$) Littrow 
mounted gratings 
(the long rectangular optics in Fig. 2);
a truly Littrow mounted grating spectrometer would have the same entrance and
exit optical paths, but in the FIFI LS system the entrance and exit paths
are slightly separated so that the outgoing beam can easily
be re-imaged onto the detectors.
This arrangement
allows for the dual use of the two anamorphic collimators 
(the two large mirrors right-most and the two large mirrors on 
top and bottom in Fig. 2).  From the image slicer, the beam
continues in each channel via the anamorphic
optics, which expand the beam to an elliptical shaped cross-section that
illuminates the grating over its 300 mm length, attaining the proposed
spectral resolution of about 1000-2000 ($\Delta \nu \sim $ 100 - 250 km/s)
In each return path, the anamorphic re-imaging is used to 
match the spatial and spectral resolution of the system to the square
pixels of the detector.

In order to cover the wide wavelength range necessary,
the Littrow-mounted gratings are operated in the 1st order
(long wavelengths) or 1st and 2nd orders (short wavelengths).
To select first or second diffraction order, exchangeable filters are
utilized in the short-wavelength branch of the spectrometer.  
The observing wavelength of each spectrograph is then tuned by tilting the
grating, and exchangeable filters are used to select the appropriate
working order of the grating.

Since the grating efficiency strongly depends upon wavelength and
diffraction order, the groove profile and separation for each grating
were optimized separately. Rigorous vectorial diffraction analysis of
the grating efficiency in the actual Littrow configuration were
performed using the PCGrate-1E Ver. 3.0 software.
The calculated grating efficiencies as a function of
wavelength and diffraction order are shown in Fig. 4. 
Our goal was to maximize the efficiencies across the two bands.
For the long wavelengths, this was done by using a symmetric grating
profile that provided a high efficiency across the band, and 
for the shorter wavelengths one grating order was insufficient, and 
the best result was obtained by maximizing the efficency over two orders
by using an asymmetric grating profile.
In general, the optical parameters of FIFI LS are summarized in Table 1.

\begin{table}
\begin{center} 
\begin{tabular}{lcc}
\hline
\rule[-1ex]{0pt}{3.5ex}
       &  Short $\lambda$ Spectrometer     &  Long $\lambda$ Spectrometer\\
\hline
\rule[-1ex]{0pt}{3.5ex}
Wavelength Range   
       &  42 - 110 $\mu$m                  &   110 - 210 $\mu$m\\
\rule[-1ex]{0pt}{3.5ex}
Field of View
       &  35$\arcsec \times$ 35$\arcsec$   & 70$\arcsec \times$ 70$\arcsec$\\
\rule[-1ex]{0pt}{3.5ex}
Num. of Spatial Pix.
       &  5$\times$5                       & 5$\times$5 \\
\rule[-1ex]{0pt}{3.5ex}
Num. of Spectral Pix.
       & 16                                &  16 \\
\rule[-1ex]{0pt}{3.5ex}
Pixel size
       & 3.6$\times$3.6  mm                & 3.6$\times$3.6 mm\\
\rule[-1ex]{0pt}{3.5ex}
f/D at collimator
       & 20                                & 10\\
\rule[-1ex]{0pt}{3.5ex}
f/D on detec. array
       & 48                                & 24\\
\rule[-1ex]{0pt}{3.5ex}
Collim. beam dia.
       &  40$\times$80 mm                  & 80$\times$160 mm\\
\rule[-1ex]{0pt}{3.5ex}
Grating lines
       &  12/mm                            & 8.5/mm\\
\rule[-1ex]{0pt}{3.5ex}
Grating angle
       &  28 - 68$\arcdeg$                 &  28 - 68$\arcdeg$\\
\rule[-1ex]{0pt}{3.5ex}
Groove profile
       & asymmetric                        & symmetric\\
\rule[-1ex]{0pt}{3.5ex}
Groove separation
      & 83 $\mu$m                          & 118 $\mu$m\\
\rule[-1ex]{0pt}{3.5ex}
Groove depth
      & 42 $\mu$m                          & 140 $\mu$m\\
\rule[-1ex]{0pt}{3.5ex}
Enclosed angle
      & 84$\arcdeg$                        & 44$\arcdeg$\\
\rule[-1ex]{0pt}{3.5ex}
Resolution (c$\Delta\lambda$/$\lambda$)
       & 100 -250 km/s                     &  100 -250 km/s\\
\rule[-1ex]{0pt}{3.5ex}
Instantaneous Vel. coverage
       & 1300 - 3000 km/s                  & 1300 - 3000 km/s\\
\hline 
\end{tabular}
\end{center}
\caption{Optical Parameters of FIFI LS.}
\end{table}

\begin{figure}
\begin{center}
\begin{tabular}{c}
\end{tabular}
\end{center}
\caption{
Efficiency of the gratings as a function of wavelength and 
diffraction order.  The insets show the grating
profiles: short $\lambda$ grating on the left
and long $\lambda$ grating on the right (not to scale).
}
\end{figure}

\subsection{Grating Mechanical Layout}

To cover the specified wavelength range of 42 to 110 $\mu$m and 110 to
210 $\mu$m in first and second order, both gratings have to be tilted by
an angle of $\pm$20$\arcdeg$. However, to reach a spectral accuracy of
$\sim$100 - 250 km/s, the grating has to be moved and controlled with
a precision of less than 4$\arcsec$ while maintaining
mechanical stability of the optical surface.
In addition, this precision must
be reached at liquid helium temperatures. In order to minimize 
deformation to the grating surface from vibration, the
grating structure design was extensively tested with finite element 
modeling.
The final grating structure, as shown in Fig. 5, has $<$ 1 $\mu$m of
deflection with 10 N of applied force at the outer edge of the structure.

\begin{figure}
\begin{center}
\begin{tabular}{c}
\end{tabular}
\end{center}
\caption{
3D model of the diffraction grating.}
\end{figure}

To obtain a large tilting range and precise positioning, 
the grating is actuated by a two stage tilting mechanism.
The first stage, for coarse positioning, consists of a support structure
connected to the bottom of the grating.
The support structure is driven by a roller-screw lever arm
mechanism (a so-called sine-bar mechanism) that tilts the grating 
via a push-pull movement.
The roller screw is driven by a stepper motor at liquid
nitrogen temperature that is thermally isolated from the grating 
by a magnetic feed-thru.
The second stage, for fine positioning, utilizes
a PZT, which drives the grating with respect to the support structure
via a directly attached lever arm. The PZT movement is controlled in a
software loop with closed-loop bandwidth up to 1kHz.  Additionally, an
eddy current damping system is mounted to the grating to minimize
in-flight vibrations of the airplane, which are a perpetual source of
error on airborne experiments, especially affecting moving parts where
resonant modes can raise error motion to an unacceptable value.

\subsection{Grating Position Read-out}

For highly reliable measurement of the angular position of the gratings, 
we directly attached an INDUCTOSYN$^{\em TM}$ position transducer.  
An INDUCTOSYN$^{\em TM}$ is effectively a transformer with the
primary and secondary windings placed on a rotor and stator,
respectively.
The winding pattern on the rotor is excited by a 10 kHz signal,
while on the stator, there are two periodic patterns that are
90$\arcdeg$ out of phase with each other.
In operation, the rotor and stator windings inductively couple
such that the two output signals from the stator have amplitudes
which vary as the sine and cosine functions based on the
relative position in each winding cycle.
By comparing the two amplitudes, a 
high-resolution difference positional accuracy is obtained.

The position transducers work with a fairly high operating current of
$\sim$ 0.25 Amperes. To provide thermal insulation of the 4K worksurface,
only thin signal wires are tolerated within the cryogenic regions. Thin
wires, on the other hand, raise the power dissipation, leading again to
higher heat input. To overcome this problem, we use a superconducting
transformer which steps down the signal voltage by a factor of 50
after entering the 4K region. In this way, a high-voltage, but low-
current excitation signal, can be conveyed via thin wires, reducing
dissipation losses.

The relative amplitudes of the two transducer outputs are a measure
of the actual position of the grating, with proper
initial calibration. Both signals are amplified
separately in a low-noise amplifier stage and translated into 14 bit
digital position data by a Resolver-to-Digital Converter. Since the
transducer output produces a roll over of the position data every 
1.4$\arcdeg$,
an additional loop counter keeps track on the position data over the
full 40$\arcdeg$ tilting range of the grating. First tests with a prototype
setup at liquid helium temperature, showed that an angular resolution
of less than 0.5$\arcsec$ can be obtained easily.

\subsection{Detectors}

As mentioned above, FIFI LS uses two 16 $\times$ 25 detector
arrays to cover the 42 - 110 $\mu$m and 110 - 210 $\mu$m
wavelength bands\cite{dirk}.
We chose to use Gallium-doped Germanium photoconductor
detectors since
they are proven to be very sensitive in the wavelength
range 40 - 120 $\mu$m, and, with the application
of $\sim$ 600 N~mm$^{-2}$ of stress, their wavelength sensitivity
shifts to 100 - 220 $\mu$m\cite{dirk,gordon,jeff}.
Thus, FIFI LS uses two Ge:Ga detector arrays, one stressed and
one unstressed; our design and initial testing
is discussed in detail in this volume\cite{dirk}.

\section{CRYOSTAT}

\begin{figure}
\begin{center}
\begin{tabular}{c}
\end{tabular}
\end{center}
\caption{
Solid model of the FIFI LS system.  FIFI LS
is mounted to the SOFIA instrument flange on the right side.
This image shows the three cryogenic containers (top-most), the
optical worksurfaces (hanging from the cryogenic containers via
G-10 tabs, colored black), the optics, and the guiding camera (bottom-most).
}
\end{figure}

The FIFI LS instrument is directly attached to the science instrument
mounting flange of the SOFIA telescope. In Fig. 6, a 3D rendered
model of the FIFI LS instrument is shown.
The cryostat is enclosed by a vacuum vessel which also provides
the mechanical interface for mounting.
Mounted to the bottom of the vacuum vessel are the
dichroic beam splitter and the field optics for the focal plane guiding
camera.
The dichroic filter separates
the telescope's light (entering Fig. 6 on the bottom right) into two beams:
the optical, directed downward to the telescope guiding camera, 
and the infrared,
directed upward into the cryostat. 
The infrared beam from the telescope enters the vacuum vessel
through a polyethylene window, which also serves as a pressure barrier
between stratospheric pressure and the vacuum inside the instrument.

\begin{figure}
\begin{center}
\begin{tabular}{c}
\end{tabular}
\end{center}
\caption{
Solid model of the three cryogenic containers,
from top to bottom: liquid nitrogen, liquid helium, and
the smaller superfluid helium.
}
\end{figure}

There are three cryogenic containers in FIFI LS for liquid
nitrogen, liquid helium, and superfluid helium.
The manner in which the three vessels are mounted together
is shown in Fig. 7, a center cut-thru the cryogenic vessels.
The liquid nitrogen container has a capacity
of 25 liters, which provides cooling for the outer radiation shields, 
the liquid nitrogen worksurface and the entrance optics 
(i.e. the K-mirror assembly and the re-imaging optics).
The cut-thru of the liquid nitrogen vessel in Fig. 7 shows two of its
inner support ribs.
The liquid nitrogen worksurface is suspended
from the warm vacuum vessel by G-10 fiberglass stand-offs which provide
high mechanical stiffness and low thermal conductivity. The liquid
nitrogen container is designed for a cryogen holding time of about
28 hours.

The 35 liter main liquid helium reservoir provides cooling for the
inner radiation shields and the liquid helium optical bench. 
The liquid helium optical bench,
suspended from the liquid nitrogen worksurface by G-10 tabs, 
mechanically supports and cools all of the optical
components after the calibration system (excluding the detectors).
The expected cryogen holding time for the main liquid helium 
reservoir is up to 50 hours.

Since the detector arrays require operating temperatures 
below 4K\cite{dirk}, they
are mounted to a small 2.8 superfluid helium tank, which is again
suspended by G-10 tabs from the liquid helium optical bench. This tank
is pumped in order to reach a temperature of $\sim$ 2K. The expected
maximum holding time for the liquid helium tank is about 18 hours.
To ensure safe operation during flight, all cryogen vessels are provided
with coaxial neck tubes and warm pressure relief valves.

\begin{figure}
\begin{center}
\begin{tabular}{c}
\end{tabular}
\end{center}
\caption{
Finite element analysis of the vacuum vessel, LN$_{2}$ vessel, and the LHe vessel
under 3.5 bars.  The distortion is overemphasized, never exceeding
1 mm of displacement.
}
\end{figure}

\subsection{Cryostat Analysis}

In order to comply with Federal Aviation Administration requirements, we are
designing the Cryostat to withstand 3 times the operational pressure---
about 3.5 bars.  Fig. 8 shows finite element analysis 
results for the three large vessels at 3.5 bars of internal
pressure-- overemphasizing the deformation for display purposes
The maximum deformation in the three vessels are
0.91 mm for the vacuum vessel, 0.20 mm for the LN$_{2}$ vessel, and 0.24 mm
for the LHe vessel.
In addition, we used FEA for determining the best
design for stability of the internal optical work surfaces
and to minimize the amount of stress on the
welding joints of the vessels.
In the latter case, for example, 
we find that the best approach is to create a nearby zone that
can elastically deform, thereby relocating areas of high stress
away from the welding seam.

\section{Scientific Capabilities}

FIFI LS employs two fixed pixel sizes of 7$\arcsec$ (short wavelength
spectrometer) and 14$\arcsec$ (long wavelength spectrometer), respectively,
determined by the image slicer. The 5 $\times$ 5 pixel fields of view are
observed simultaneously with two Ge:Ga photoconductor arrays. Observing
wavelength are adjusted by tilting the Littrow mounted grating in each
channel. Spectral coverage of $\sim$ 1500 km/s around a selected far-infrared
line is obtained simultaneously for all 25 spatial pixels. A summary of
important instrument properties is shown in Table 2.

\begin{table}
\begin{center}
\begin{tabular}{llcc}
\hline
\rule[-1ex]{0pt}{3.5ex}
Wavelength Range&
       &  42 - 110 $\mu$m                 &  110 - 210 $\mu$m\\
\hline
\rule[-1ex]{0pt}{3.5ex}
Pixel size&
       &  7$\arcsec \times$ 7$\arcsec$    &   14$\arcsec \times$ 14$\arcsec$\\
\rule[-1ex]{0pt}{3.5ex}
Field of view 5$\times$5 pixels&
       &  35$\arcsec \times$ 35$\arcsec$  &   70$\arcsec \times$ 70$\arcsec$\\
\rule[-1ex]{0pt}{3.5ex}
Resolution (c$\Delta\lambda$/$\lambda$)&
       & 100 -250 km/s                     &  100 -250 km/s\\
\rule[-1ex]{0pt}{3.5ex}
Instantaneous Vel. coverage&
       & 1300 - 3000 km/s                  & 1300 - 3000 km/s\\
\rule[-1ex]{0pt}{3.5ex}
Point source det. limit & $\lambda$~=~50$\mu$m
       & 5.5$\times 10^{-17}$ W/m$^{2}$    & ..... \\
\rule[-1ex]{0pt}{3.5ex}
~~(5$\sigma$ in 1 hr)     & $\lambda$~=~100$\mu$m
       & 3.5$\times 10^{-17}$ W/m$^{2}$    & ..... \\
\rule[-1ex]{0pt}{3.5ex}
                        & $\lambda$~=~150$\mu$m
       & .....                             & 2.2$\times 10^{-17}$ W/m$^{2}$\\
\rule[-1ex]{0pt}{3.5ex}
                        & $\lambda$~=~200$\mu$m
       & .....                             & 1.4$\times 10^{-17}$ W/m$^{2}$\\
\hline 
\end{tabular}
\end{center}
\caption{FIFI LS performance specifications.}
\end{table}

 

  \end{document}